\documentclass{iopart}

\usepackage{graphicx}
\usepackage{cite}
\usepackage{iopams}

\newcommand{\ep}{\epsilon}

\newcommand{\bea}{\begin{eqnarray}}
\newcommand{\beq}{\begin{equation}}
\newcommand{\eea}{\end{eqnarray}}
\newcommand{\eeq}{\end{equation}}

\begin{document}

\title[Spectral singularities in $\mathcal{PT}$-symmetric Bose-Einstein
condensates]{Spectral singularities in $\mathcal{PT}$-symmetric Bose-Einstein
  condensates}

\author{W D Heiss$^{1, 2}$, H Cartarius$^3$, G Wunner$^{1, 3}$, J Main$^3$}
\address{$^1$Department of Physics, University of Stellenbosch,
  7602 Matieland, South Africa}
\address{$^2$National Institute for Theoretical Physics (NITheP), Western Cape,
  South Africa}
\address{$^3$Institut f\"ur Theoretische Physik 1, Universit\"at Stuttgart,
  Pfaffenwaldring 57, 70569 Stuttgart, Germany}
\ead{Holger.Cartarius@itp1.uni-stuttgart.de}

\begin{abstract}
  We consider the model of a $\mathcal{PT}$-symmetric Bose-Einstein condensate
  in a delta-functions double-well potential. We demonstrate that analytic
  continuation of the primarily non-analytic term $|\psi|^2\psi$ -- occurring
  in the underlying Gross-Pitaevskii equation -- yields new branch points
  where three levels coalesce. 
  We show numerically that the new branch points
  exhibit the behaviour of exceptional points of second and third order. A
  matrix model which confirms the numerical findings in analytic terms is
  given. 
\end{abstract}

\submitto{\JPA}

\maketitle

\section{\label{sec:intro} Introduction}
It is a characteristic feature of non-Hermitian but $\mathcal{PT}$-symmetric
Hamiltonians that, as an external parameter is varied, pairs of real
eigenvalues coalesce at an exceptional point (EP) and turn into complex
conjugate pairs (Bender and Boettcher \cite{Bender98}, Moiseyev
\cite{Moiseyev2011a}).
Hamiltonians are called $\mathcal{PT}$ symmetric if they commute with
the combined action of the parity ($\mathcal{P}$: $x \to -x$, $p \to -p$)
and time reversal ($\mathcal{T}$: $x \to x$, $p \to -p$, $\mathrm{i}
\to -\mathrm{i}$) operators, i.e.\ $[\mathcal{PT},H] = 0$.
For non-Hermitian Hamiltonians in position space the condition for
being $\mathcal{PT}$ symmetric is $V(-x) = V^{\ast}(x)$, i.e.\ the real
and imaginary part of the potential has to be a symmetric and antisymmetric
function of $x$, respectively. Imaginary potentials are used to model effects
of gain (positive imaginary part: source term) and loss (negative
imaginary part: sink term), i.e.\ an increase or a decrease of the
probability density. For Bose-Einstein condensates gain (loss) of the
probability density is achieved by coherently adding (removing) particles to
the condensate. The gain and loss effects provide access to interesting physical
properties as such non-Hermitian Hamiltonians can have real eigenvalue spectra
\cite{Bender98}. Dissipation effects may also be exploited to enhance
coherence properties of quantum systems
\cite{Diehl2008a,Witthaut2008b,Krauter2011a}, which can be important in
quantum information processing.

Recent investigations of {\em non-linear} $\mathcal{PT}$-symmetric Hamiltonians
modelling Bose-Einstein condensates in double-well potentials with sink and
source terms of atoms at the respective wells have revealed a specific spectral
behaviour which implies apparently new spectral singularities
\cite{Graefe08a,Graefe08b,Graefe10,Graefe12b,Cartarius12b,Cartarius12c,Dast13a}.
Pairs of real eigenvalues merge at an EP when the strength of the
non-Hermiticity  is increased, but no pairs of complex conjugate 
eigenvalues appear beyond the EP. Rather, such pairs are born as if ``out of
nowhere'' {\em before} the EP,  that is for smaller values of the loss/gain
term, where they bifurcate from the branch of the ground state eigenvalue. 

The reason for this unusual behaviour is the non-analyticity of the non-linear
term $|\psi|^2\psi$ in the underlying Gross-Pitaevskii equation.
Numerical calculations for Bose-Einstein condensates in $\mathcal{PT}$-symmetric
double-well potentials have shown \cite{Cartarius12b, Cartarius12c, Dast13a}
that by appropriate analytic continuation of the non-linear term 
pairs of complex conjugate eigenstates indeed arise at the branch point.
Moreover, in an analytically solvable toy model for a $\mathcal{PT}$-symmetric
two-mode Bose-Einstein condensate Graefe \cite{Graefe12b} used a particular
analytic extension of eigenvalues and eigenstates. This way it has been
demonstrated that the ``prematurely'' born pairs of complex conjugate
eigenvalues emerge from an additional pair of real eigenvalues. Therefore three
real eigenvalues effectively coalesce at what we shall term a triple point. 

The features of a non-linear term of the form $|\psi|^2\psi$ have become very
relevant for the study of $\mathcal{PT}$-symmetric physical systems since this
type of non-linearity not only appears in the mean-field description of
Bose-Einstein condensates. $\mathcal{PT}$-symmetric optical setups of coupled
dual waveguides including a Kerr non-linearity show exactly the same structure
in the underlying equations. They have been used to demonstrate the existence
of uni-directional structures \cite{Ramezani10} or the presence of solitons
in loss/gain media \cite{Musslimani2008a,Abdullaev10,Bludov10,Driben2011a,%
  Abdullaev2011,Bludov2013a}. The remarkable success of realising
$\mathcal{PT}$ symmetry and $\mathcal{PT}$-symmetry breaking experimentally
emphasises the relevance of these optical setups \cite{Guo09,Rueter10}. The
additional properties of the combination with a non-linearity as the
possibility of uni-directionality \cite{Ramezani10} might lead to new technical
devices. In quantum mechanics non-linear $\mathcal{PT}$-symmetric systems have
been discussed in model potentials by Musslimani et al.\
\cite{Musslimani2008b}, and for Bose-Einstein condensates described in a
two-mode approximation \cite{Graefe08a,Graefe08b,Graefe10,Graefe12b} or by
solving the mean-field limit of the Gross-Pitaevskii equation in position space
\cite{Cartarius12b,Cartarius12c,Dast13a}.

Potentials consisting only of delta-functions often provide a very simple
access to solutions, in many cases they can be obtained analytically. For
example, the spontaneous symmetry breaking in double-well structures has been
investigated analytically in the limit of infinitely narrow potential wells
described by delta-functions by  Mayteevarunyoo et al.\
\cite{Mayteevarunyoo2008a}. Rapedius and Korsch investigated the decay of
Bose-Einstein condensates in a double-delta-setup \cite{Rapedius2009a}.
Bifurcations of stationary solutions have been studied by Witthaut et al.
\cite{Witthaut2008a} for Bose-Einstein condensates in a delta-comb. With the
exception of the study of Bose-Einstein condensates in a
$\mathcal{PT}$-symmetric double-delta-trap \cite{Cartarius12b,Cartarius12c},
the discussion of $\mathcal{PT}$-symmetry in delta-potentials has been
restricted to linear quantum mechanics. 
Jakubsk\'y and Znojil \cite{Jakubsky05} considered analytical solutions of a
particle in an infinitely high square well containing two purely imaginary
delta-functions in a $\mathcal{PT}$-symmetric arrangement. The influence of
a pair of $\mathcal{PT}$-symmetric delta-functions on the bound and scattering
states of a single real delta-function has been investigated by Jones
\cite{Jones2008a}. The spectral properties, in particular bound states and
spectral singularities, in non-Hermitian delta-potentials have been studied by
Mostafazadeh et al.
\cite{Mostafazadeh2006a,Mostafazadeh2009a,Mehri-Dehnavi2010a}.

It is the purpose of the present paper to study the new types of spectral
singularities in more detail. We do this in Section \ref{sec:GPE} by solving
numerically an analytically continued version of the Gross-Pitaevskii equation
for the model of a Bose-Einstein condensate in a $\mathcal{PT}$-symmetric double
delta potential.  We find that the triple point can exhibit the
behaviour of a second-order (EP2) or a third-order (EP3) exceptional point
depending on the particular encirclement. To achieve this an appropriate
analytic extension of the Gross-Pitaevskii equation is used to reveal the
mathematical structure of the branch points. In a matrix model presented in
Section 3 this behaviour is described in analytic terms, by which
more insight is gained into the behaviour of the spectral singularities, and it 
allows for a detailed analysis of the nontrivial limit of a vanishing
nonlinearity in the Gross-Pitaevskii equation.

\section{Complexification of the Gross-Pitaevskii equation}
\label{sec:GPE}

\subsection{Model with delta-function traps}

The model of a Bose-Einstein condensate trapped in two $\mathcal{PT}$-symmetric
potential wells represented by delta-functions has been introduced in
\cite{Cartarius12b,Cartarius12c}. Here we only recall the essential points
which are relevant to the present discussion. In the mean-field description
the non-linear Gross-Pitaevskii equation to be solved reads in dimensionless
units 
\begin{eqnarray}\label{eq:GPEdimless}
  -\psi^{\prime\prime}(x) &-&\left[(1+\rmi \gamma)\delta(x+a/2) 
    + (1-\rmi\gamma)\delta(x-a/2)\right]\psi(x) \nonumber \\
  &-& g|\psi(x)|^2 \psi(x)
  = -\kappa^2 \psi(x)\, .
\end{eqnarray}
Throughout the paper we fix the normalisation by $\int |\psi(x)|^2 {\rm d}x=1$.
Here $g$ is the non-linear interaction strength, the real-valued parameter
$\gamma$ determines the strength of the gain and loss terms in the two delta
wells located at $\pm a/2$. (The relation to physical quantities can be found
in \cite{Cartarius12b,Cartarius12c} as well as details on the numerics.)
The gain and loss terms have here been introduced on the level of
the Gross-Pitaevskii equation, i.e. they influence the probability density
of the whole condensed phase. The physical interpretation is a coherent
gain or loss of particles, i.e. particles are added or removed directly to
or from the condensate. A physical realisation of this process is possible
by an active coupling of the two wells to a reservoir, e.g.\ a third well in
which a condensate is trapped, or by embedding the double well in a transport
chain, a system which has shown to exhibit the same stationary states and
dynamics \cite{Kreibich2013a}. Note that the treatment of loss effects in
cold atom systems with a master equation \cite{Trimborn2008a,Witthaut2011a}
leads to a similar equation in the mean-field limit if coherence is preserved.
A slightly different equation is obtained in \cite{Graefe08b}, where the
gain and loss is introduced on the single-particle level.

\begin{figure}[tb]
  \begin{center}
    \includegraphics[width=\textwidth]{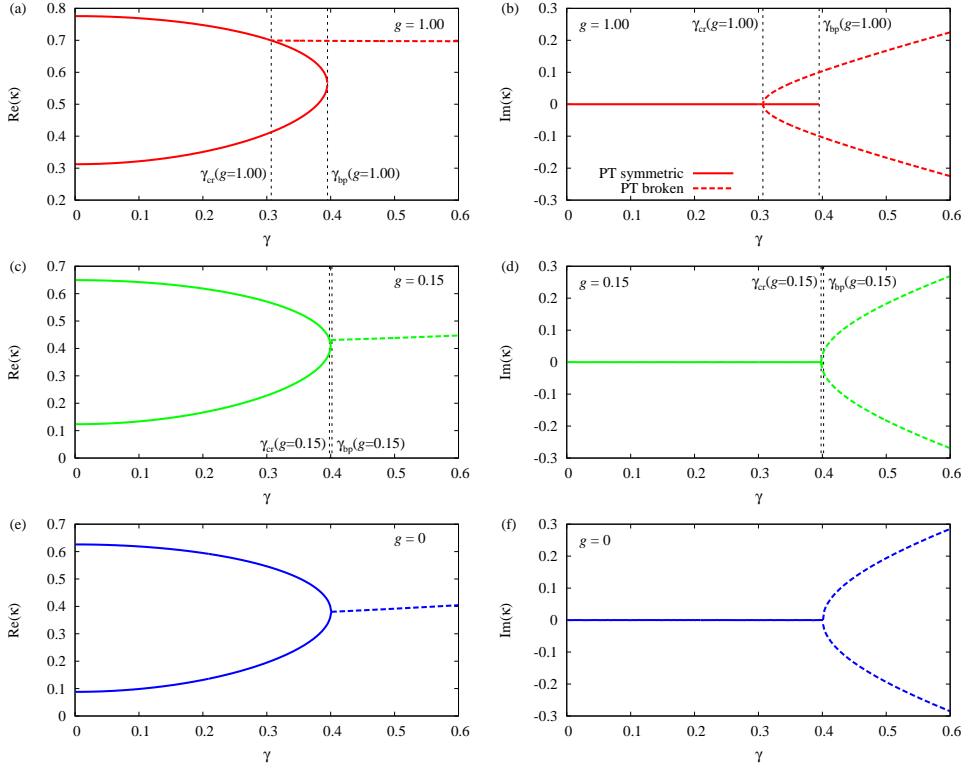}
  \end{center} 
  \caption{\label{fig:kappa_spectra_nonanalytic}
    Real parts [(a), (c), (e)] and imaginary parts [(b), (d), (f)] of the 
    eigenvalues $\kappa$ of the full non-linear equation (\ref{eq:GPEdimless})  
    as functions of the loss/gain parameter $\gamma$ for $a=2.2$
    and the three values of the non-linearity parameter $g = 1.00$
    [(a) and (b)], $0.15$ [(c) and (d)] and $0$ [(e) and (f)].
    Solid lines denote purely real eigenvalues, dashed lines
    complex conjugate eigenvalues.
    For $g\neq 0$ the complex conjugate eigenvalues bifurcate
    from the ground state branch at $\gamma_{\mathrm{cr}}$ before
    the branch point $\gamma_{\mathrm{bp}}$ where the two real
    solutions coalesce. Between $\gamma_{\mathrm{cr}}$ and 
    $\gamma_{\mathrm{bp}}$ the $\mathcal{PT}$-broken and the
    $\mathcal{PT}$-symmetric branches coexist.}
\end{figure}

We are interested in the lower end of the eigenvalue spectrum,
including complex eigenvalues, i.e.\ solutions of (\ref{eq:GPEdimless}) are
searched with $\kappa \in {\mathbb {C}}$. Throughout the article
we will only investigate bound states, i.e. square integrable wave functions
with $\psi(\pm\infty) = 0$ and no scattering solutions.
Figure~\ref{fig:kappa_spectra_nonanalytic}
recapitulates the results for the eigenvalues $\kappa$ for different
values of the non-linearity parameter $g$. It can be seen that for every value
of $g$ a pair of real eigenvalues exists up to some value $\gamma_{\mathrm{bp}}$,
where they coalesce at a branch point. Pairs of complex conjugate eigenvalues
emerge at critical values $\gamma_{\mathrm{cr}} < \gamma_{\mathrm{bp}}$, where
$\gamma_{\mathrm{cr}}$ decreases for increasing $g$.
We discuss only examples with $g \ge 0$ in this article, for which
the complex conjugate eigenvalues bifurcate from the ground state.
Note that the ground state is the upper branch in Figure 
\ref{fig:kappa_spectra_nonanalytic} since the largest value $\kappa$ leads
to the lowest energy $E = -\kappa^2$. We mention that for $g < 0$
the complex eigenvalues branch off from the excited state.
For sufficiently large non-linearity these solutions already appear for
$\gamma = 0$. Thus one has ranges of $\gamma$ where two real and two
complex conjugate eigenvalues coexist. The wave functions show that those
belonging to the real eigenvalues are themselves $\mathcal{PT}$-symmetric,
i.e.~they are eigenfunctions of $\mathcal{PT}$, whereas the two states with
complex eigenvalues $\kappa$ are not.

Figure~\ref{fig:kappa_spectra_nonanalytic} reveals that the branching-off of
the real eigenvalue changes continuously from the branch for $g = 0$. The wave
functions, however, show a non-uniform limit as $g$ tends to zero: for every
value $g > 0$ their asymptotic form is given by $1/\cosh(\kappa x)$, while at
$g = 0$ it is $\exp(\pm \kappa x)$. The non-uniform behaviour of $g\to 0$ will
be addressed in more detail in Section 3.

\begin{figure}[tb]
  \begin{center}
    \includegraphics[width=\textwidth]{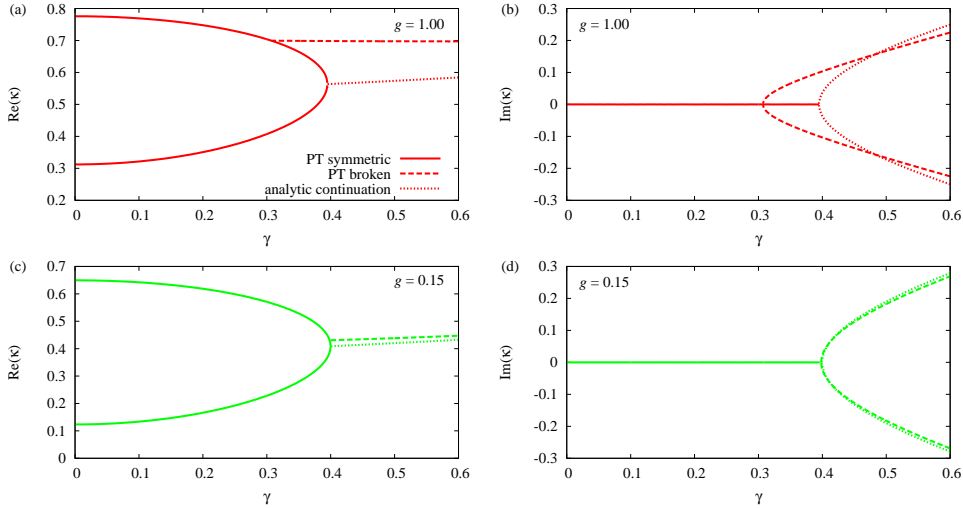}
  \end{center}
  \caption{\label{fig:kappa_spectra_PT_continuation}
    Real parts [(a) and (c)] and imaginary parts [(b) and (d)] of the
    eigenvalues $\kappa$ for the states appearing at $g = 1$ [(a) and (b)]
    and $g=0.15$ [(c) and (d)]. Using the substitution
    $|\psi(x)|^2 \to \psi(x) \psi(-x)$ the analytic continuation beyond the
    branch point $\gamma_{\mathrm{bp}}$ of the two real $\mathcal{PT}$-symmetric
    eigenvalue states is obtained.}
\end{figure}

Pairs of complex conjugate eigenvalues emerge at the branch points only when
the non-analytic term  $|\psi(x)|^2$ in (\ref{eq:GPEdimless}) is continued
beyond the branch points. This is illustrated in
figure~\ref{fig:kappa_spectra_PT_continuation}.
For the analytic continuation we exploit the $\mathcal{PT}$-symmetry of
the wave functions corresponding to the real eigenvalues, i.e.~$\psi^\ast(x)
= \psi(-x)$. Therefore, when approaching the branch point we replace the
non-linear term $g|\psi(x)|^2$ for the $\mathcal{PT}$-symmetric states by
$g\, \psi(x) \psi(-x)$. This function can be continued analytically. We note
that in the numerical calculation the additional condition $\int \psi(x)
\psi(-x) dx = 1$ must be enforced to fix the phase of the non-linearity in the
$\mathcal{PT}$ broken regime. 

The substitution of $|\psi(x)|^2$ by $\psi(x) \psi(-x)$ allowing analytic
continuation of the square modulus of $\mathcal{PT}$-symmetric eigenstates
shows that the number of solutions does not change at the branch point
$\gamma_{\mathrm{bp}}$. Yet, at $\gamma < \gamma_{\mathrm{cr}} < \gamma_{\mathrm{bp}}$
the number of states seem to be less. In fact, the pair of complex eigenvalues
bifurcating from the ground state at $\gamma_{\mathrm{cr}}$ seem to come about
without real-valued precursors. However, employing a different strategy
of complex continuation these precursors and their corresponding eigenstates
can indeed also be found.

\subsection{Analytic continuation}

The procedure \cite{Cartarius08a} is to express all complex quantities by two
real-valued functions, either amplitude and phase, or real and imaginary
parts. The non-analytic Gross-Pitaevskii equation then decomposes into two
coupled real-valued differential equations which can be continued analytically.

We write for the wave function, the double well potential and the eigenvalue 
$\psi = \psi_{\rm r} + \rmi \psi_{\rm i},\;V = V_{\rm r} + \rmi V_{\rm i}$ and
 $\kappa = \kappa_{\rm r} + \rmi \kappa_{\rm i}$, respectively.
Inserting into the Gross-Pitaevskii equation and sorting
the real and imaginary contributions leads to
\numparts
\begin{eqnarray}
-\psi_{\rm r}^{\prime\prime} + V_{\rm r}\psi_{\rm r} - V_{\rm i}\psi_{\rm i}
-g(\psi_{\rm r}^2 + \psi_{\rm i}^2) \psi_{\rm r} & =
\kappa_{\rm r} \psi_{\rm r} - \kappa_{\rm i} \psi_{\rm i} \, ,\\
-\psi_{\rm i}^{\prime\prime} + V_{\rm r}\psi_{\rm i} + V_{\rm i}\psi_{\rm r}
-g(\psi_{\rm r}^2 + \psi_{\rm i}^2) \psi_{\rm i} & =
\kappa_{\rm r} \psi_{\rm i} + \kappa_{\rm i} \psi_{\rm r} \,.
\end{eqnarray}
\endnumparts
The full analytical extension is implemented by allowing these real and
imaginary parts of the wave functions and eigenvalues to become themselves
complex:
\numparts
\begin{eqnarray} 
\psi_{\rm r} &= \psi_{{\rm r} {\rm r}} + \rmi \psi_{{\rm r} {\rm i}}\, , \quad
\psi_{\rm i} = \psi_{\rm i \rm r} + \rmi \psi_{\rm i \rm i} \, ,
\label{eq:extension1}\\
\kappa_{\rm r} & = \kappa_{\rm r \rm r} + \rmi \kappa_{\rm r \rm i}\, , \quad
\kappa_{\rm i}  = \kappa_{\rm i \rm r} + \rmi \kappa_{\rm i \rm i} \,.
\label{eq:extension2}
\end{eqnarray}
\endnumparts
This extension allows for continuing both the $\mathcal{PT}$-symmetric and the
$\mathcal{PT}$-broken states beyond the corresponding branch points.
When plotting the eigenvalues we recombine the 4 real quantities
$\kappa_{\rm r \rm r}, \kappa_{\rm r \rm i}, \kappa_{\rm i \rm r}, \kappa_{\rm i \rm i}$ 
into the real and imaginary parts of $\kappa$, i.\,e.\
\begin{equation}
\kappa = (\kappa_{\rm r \rm r} -\kappa_{\rm i \rm i}) + \rmi (\kappa_{\rm r \rm i}
+ \kappa_{\rm i \rm r})
= \mathrm{Re}(\kappa ) + \rmi \, \mathrm{Im}(\kappa )   \,.
\label{eq:combination}
\end{equation}

\begin{figure}[tb]
  \begin{center}
    \includegraphics[width=\textwidth]{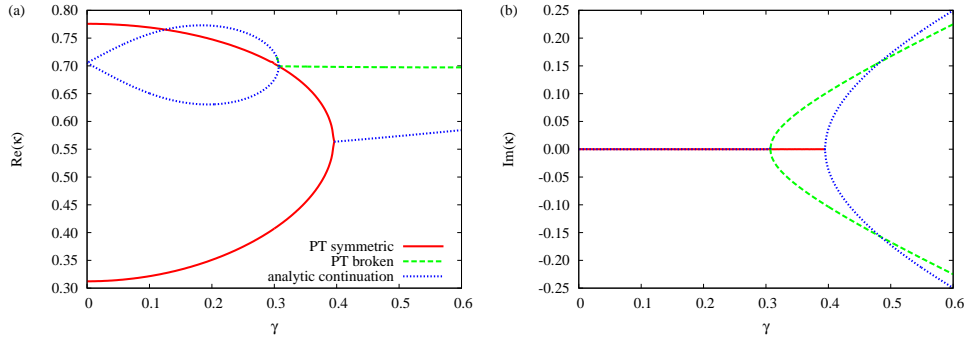}
  \end{center}
  \caption{\label{fig:g100} 
    Real parts $\mathrm{Re}(\kappa)$ (a) and imaginary parts $\mathrm{Im}
    (\kappa)$ (b) of the spectrum after the analytical extension for $g = 1$.
    The red solid and green dashed lines represent the $\mathcal{PT}$-symmetric
    and $\mathcal{PT}$-broken states already known without any analytic
    continuation. The new states visible with the full
    analytic continuation (\ref{eq:extension1}), (\ref{eq:extension2}), 
    (\ref{eq:combination}) are marked by blue dotted lines. With these
    continued states the number of states does no longer change with $\gamma$.}
\end{figure}

For the value of the non-linearity parameter $g=1$ the results for
$\mathrm{Re}(\kappa)$ and $\mathrm{Im}(\kappa)$ are shown in 
figure~\ref{fig:g100}.
For $\gamma \le \gamma_{\mathrm{cr}}$ there are indeed two new branches of
real-valued $\kappa$. They merge at the triple point and give rise to the
complex conjugate pair of eigenvalues. Moreover, these two solutions are
present even at $\gamma = 0$, where they refer to two degenerate eigenvalues.

\begin{figure}[tb]
  \begin{center}
    \includegraphics[width=0.5\textwidth]{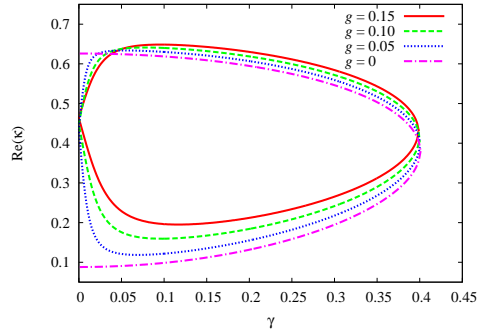}
  \end{center}
  \caption{\label{fig:small_g} 
    Eigenvalue spectra for $g = 0.15$ (red solid lines), $g=0.10$ (green
    dashed lines), and $g=0.05$ (blue dotted lines). For comparison 
    the spectrum without the non-linear term ($g=0$) is also displayed .} 
\end{figure}

It is remarkable  that the spectra of the two additional states obtained by
the analytic continuation become very similar to the spectra of the original
$\mathcal{PT}$-symmetric solutions when $g$ is decreased. This is demonstrated
in figure~\ref{fig:small_g},
where the spectra shown should be compared with those in
figure~\ref{fig:kappa_spectra_nonanalytic}. There is, however, one difference:
the two additional solutions are degenerate at $\gamma = 0$ for any $g \neq 0$,
while the eigenvalues of the two original $\mathcal{PT}$-symmetric states
always have different values at this point. 

\begin{figure}[tb]
  \begin{center}
    \includegraphics[width=\textwidth]{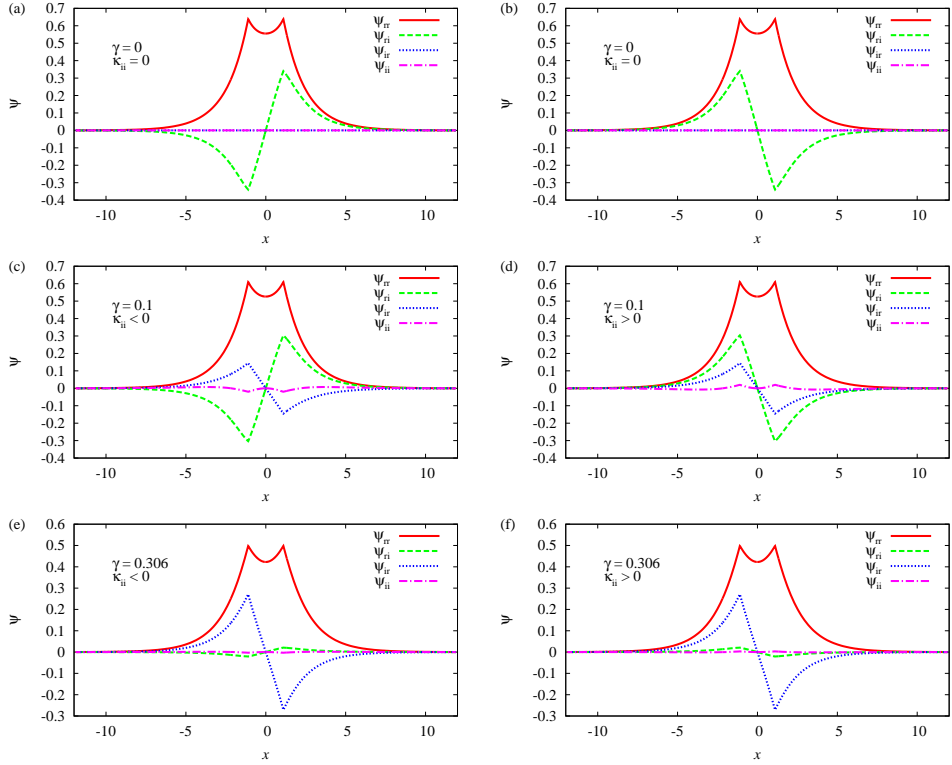}
  \end{center}
  \caption{\label{fig:wavefct} 
    Wave functions of the analytically extended solutions for a few values of
    $\gamma $ and $g=1$. The right (left) column refers to the lower (upper)
    branch of the eigenvalues.}
\end{figure}

For $g>0$ the wave functions of the analytically continued states shown in
figure~\ref{fig:small_g} differ of course from those of the original real
eigenvalues where only $\kappa_{\mathrm{rr}}$ features. It is of interest to have
a look at these wave functions. They are illustrated for $g=1$ in
figure~\ref{fig:wavefct}
for a few values of $0 \le \gamma \le  \gamma_{\mathrm{cr}}$.
Note that the functions are eigenfunctions of $\mathcal{PT}$ throughout in that
their real parts are symmetric and their imaginary parts antisymmetric. Note
further that there is a non-vanishing imaginary part $\psi_{\mathrm{ri}}$ due to
the analytic continuation even at $\gamma =0$. At this point the two wave
functions are associated with the same energy value as $\kappa_{\rm ii}=0$,
i.e.~there is a genuine degeneracy. (This in in contrast to the wave functions
of the other two energies appearing for $\gamma =0$ (see
figure~\ref{fig:g100})). The imaginary part  $\psi_{\mathrm{ri}}$ tends to zero
when $ \gamma_{\mathrm{cr}}$ is approached while another imaginary part
$\psi_{\mathrm{ir}}$ -- not due to the analytical continuation -- emerges.
Eventually, that is at  $\gamma_{\mathrm{cr}}$, the two wave functions become
equal to themselves and to the third (original) state with which they coalesce
at the triple point. From this point onwards the states are found without
the complex extension, in fact the parts invoked from the extension vanish.
Note also, that the additional real part $\psi_{\rm ii}$ switches on and off
at the endpoints of the interval $[0,\gamma_{\mathrm{cr}}$] similar to the
behaviour of $\kappa_{\rm ii}$.

\subsection{Exceptional point behaviour}

As pointed out above the three eigenfunctions are identical at the triple
point. Therefore the question arises whether or not the triple point is a
third-order exceptional point (denoted by EP3).
This is a crucial difference between the nonlinear Gross-Pitaevskii
equation \cite{Graefe12b,Cartarius12b,Cartarius12c,Dast13a} and linear
$\mathcal{PT}$-symmetric quantum systems, in which the additional triple
point does not appear.
Exceptional points are isolated singularities of the spectrum
in the physical parameter space. We denote by EP2 the point where two
states coalesce, i.e.\ their eigenvalues \emph{and} their wave functions
are identical; similarly, an EP3 denotes the point where three states
coalesce. As a function of the parameter the levels and state vectors have a
square root singularity at an EP2 and a cubic root singularity at an EP3.
As a consequence the eigenvalues and the wave functions are interchanged at
an EP2 (up to a possible phase of the latter, see e.g.~\cite{WDH2012}) when
the exceptional point is encircled on a closed contour in parameter space.
The three identical wave functions at the triple point indicate an
EP3, i.e. a cubic root branch point. For a closed contour around this point
a cyclic permutation of all three eigenvalues and eigenfunctions is then
observable (see e.g.~\cite{Graefe12a}).

\begin{figure}[tb]
  \begin{center}
    \includegraphics[width=0.5\textwidth]{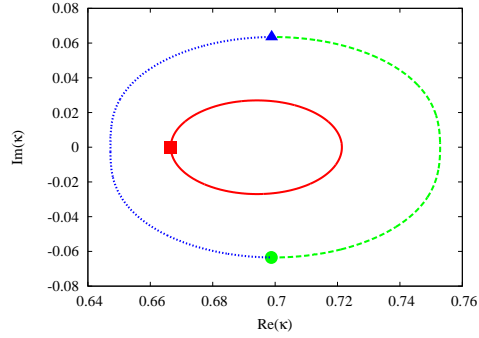}
  \end{center}
  \caption{\label{fig:gamma_circle}
    Exceptional point behaviour: circle with complex $\gamma = 0.308 + 0.04
    \mathrm{e}^{\mathrm{i} \varphi}$, $\varphi = 0\dots 2\pi$, for $g=1$, where 
    $\gamma \approx 0.308$ is the triple point. For $\varphi = 0$ we have
    $\gamma = 0.312$ and the three states exist without complex continuation.
    The red square marks the ground state with real $\kappa$. Its trace on
    the parameter space circle is represented by the red solid line. The
    two states with positive and negative imaginary parts of the eigenvalues
    $\kappa$ are represented by the blue triangle (blue dotted line for
    $\varphi \neq 0$) and the green circle (green dashed line), respectively.
  }
\end{figure}

To test numerically the precise character of the exceptional
point one parameter must be extended into the complex plane. Confirmation is
obtained by encircling the point of coalescence and checking the appropriate
change of the eigenvalues and state vectors.
Figure~\ref{fig:gamma_circle}
shows the result for a non-linearity of $g = 1$ when the triple point at
$\gamma_{\mathrm{cr}} \approx 0.308$ is encircled in the complex extended
$\gamma$ plane. One clearly sees that only the two complex eigenvalues
interchange while the ground state remains unaffected. The same holds for the
corresponding eigenfunctions. It thus appears that we deal with an EP2
corresponding to the merging of the two new eigenvalues.  

\begin{figure}[tb]
  \begin{center}
    \includegraphics[width=0.5\textwidth]{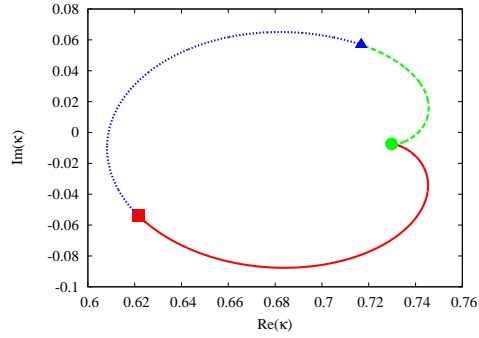}
  \end{center}
  \caption{\label{fig:asym_circle}
    Exceptional point behaviour: circle with complex $A = 0.04
    \mathrm{e}^{\mathrm{i} \varphi}$, $\varphi = 0\dots 2\pi$, for $g=1$
    and $\gamma = 0.308$. The triple point is located exactly at $A=0$. The
    symbols and lines have the same meaning as in Fig.\ \ref{fig:gamma_circle}.
    The circle in the asymmetry parameter clearly reveals the nature of an EP3.
  }
\end{figure}

However, when we allow an arbitrarily small asymmetry for the real part of the
double well potential by considering $V_{\rm asym} = A \left[ \delta(x-a/2)
- \delta(x+a/2) \right]$, and encircle now the branch point around $A = 0$ in
the complex extended $A$ plane, figure~\ref{fig:asym_circle}
clearly points to an EP3. We stress that any other perturbation would produce
the same result: the apparent EP2 which does not involve the third state becomes
a true EP3 nesting together all three states as soon as the problem becomes
disturbed, no matter of what nature and how small the disturbance may be.
This behaviour is similar to that of a study of a simple matrix model by
Demange and Graefe \cite{Graefe12a} where three coalescing eigenvectors can
exhibit both the behaviour of second- or third-order exceptional point
depending on the choice of parameters. The same behaviour has also been found 
in the investigation of pitchfork bifurcations occurring in Bose-Einstein
condensates with dipolar interaction \cite{Gut2013}. 

To summarise: the numerical solution of the $\mathcal{PT}$-symmetric
Gross-Pitaevskii equation (\ref{eq:GPEdimless}) brings about spectral
properties calling for deeper insight of the physical system. In the
following section the precise nature of these spectral singularities
will be clarified in analytic terms by means of a matrix model
simulating the properties of the non-linear physical system.

\section{The Matrix Model}
Here we present a simple low-dimensional matrix model that simulates the 
parameter dependence of the spectrum of the numerical solutions of the
Gross-Pitaevskii equation as discussed in the previous section. In particular,
the various types of the EPs as found above are of crucial importance: the
simulation must reflect the particular types of these spectral singularities.

\subsection{General properties}
\begin{figure}[tb]
  \begin{center}
    \includegraphics[width=\textwidth]{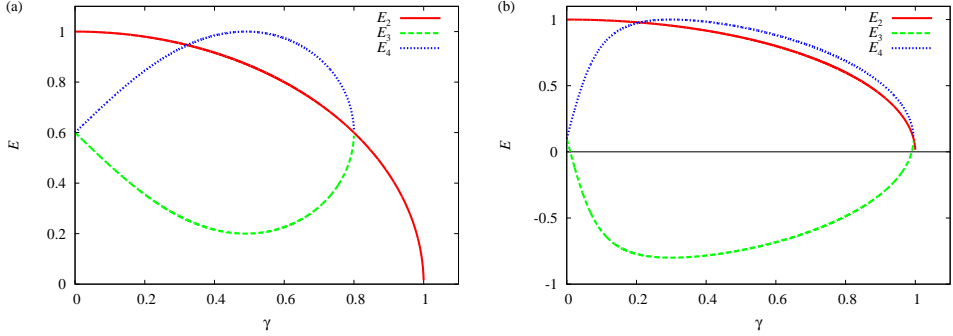}
  \end{center}
  \caption{\label{fig:spectra} 
    Eigenvalues $E_2$, $E_3$ and $E_4$ from Eq.~(\ref{eq:eigenvalues}) as a
    function of the non-Hermiticity parameter $\gamma$ for different strengths
    of the non-linearity, viz.\ $g = 1.2$ (a), $g= 0.2$ (b). Note the similarity
    with the spectra in figure~\ref{fig:small_g}.}
\end{figure}

Our starting point is a two-mode model for a $\mathcal{PT}$-symmetric
Bose-Einstein condensate \cite{Graefe12b} that mimics well that of the full
Gross-Pitaevskii equation. In the two-mode approximation the
$\mathcal{PT}$-symmetric stationary GP-equation assumes the form
\beq
\pmatrix{-\rmi \gamma + g |\phi_1|^2 & 1 \cr 1  & + \rmi \gamma + g |\phi_2|^2}
\pmatrix{\phi_1 \cr \phi_2}
= E \pmatrix{\phi_1 \cr \phi_2}
\label{eq:GPE_2_mode}
\eeq
with the normalisation $|\phi_1|^2 + |\phi_2|^2 = 1$.
The off-diagonal element in the Hamiltonian describes the tunnelling between
the two modes, and has been normalised to unity. The analytic extension of the
GP-equation in the spirit of Ref.~\cite{Cartarius08a} then yields the
eigenvalues \cite{Graefe12b}
\bea
E_1&=& -\sqrt{1-\gamma^2} \,,\nonumber \\
E_2&=& +\sqrt{1-\gamma^2} \,, \nonumber \\
E_3&=& g/2-\gamma \sqrt{\frac {1-\gamma^2-g^2/4}{\gamma^2+g^2/4}} \,, 
\nonumber \\
E_4&=& g/2+\gamma \sqrt{\frac {1-\gamma^2-g^2/4}{\gamma^2+g^2/4}} \, .
\label{eq:eigenvalues}
\eea
Note that here $\gamma_{\rm bp}=1$ and $\gamma_{\rm cr}=\sqrt{1-g^2/4}$.
In figure~\ref{fig:spectra} such spectra are illustrated for
some values of $g$. It is striking how well these eigenvalues
quantitatively describe the behaviour of those found numerically from the
GP-equation. This perfect simulation includes in particular the spectral
singularities and also allows to study the precise behaviour of the
non-uniform limit $g\to 0$.

At the critical point $\gamma=\sqrt{1-g^2/4}$ the three eigenvalues $E_2,E_3$
and $E_4$ all assume the same value $g/2$. Beyond this point $E_3$ and $E_4$
become complex; in fact, $E_3$ and $E_4$ have a square root singularity that
appears to have some characteristics of a usual EP. The precise nature,
especially its interplay with $E_2$, cannot be deduced from the spectrum alone.
It is the interaction giving rise to particular wave functions that determines
the precise nature of a spectral singularity. As we are interested in the
intersection point of the three levels  $E_2,E_3$ and $E_4$ we restrict
ourselves to a three-dimensional matrix model and leave out the level $E_1$
altogether as it seems immaterial to our specific interest here.

We seek a three-dimensional matrix, i.e.~a model Hamiltonian which produces
the spectrum $\{E_2,E_3,E_4\}$, but in such a way that all levels are fully
interacting. This is in contrast to the matrix given in \cite{Graefe12b} where
an interaction between the two branches $E_1,E_2$ and $E_3,E_4$ is not taken
into account. In other words, we seek a similarity transformation $s$ such that
\beq
ham=s\cdot j\cdot s^{-1}
\eeq
where $j$ is diagonal with the levels $\{E_2,E_3,E_4\}$ in its diagonal and
$s$ invokes the interaction between the levels. For our purpose the matrix $s$
must bear all singularities of the eigenvalues, that is of $j$. Moreover, 
a particular rank drop of $s$ at the
critical point $\gamma=\sqrt{1-g^2/4}$ indicates as to whether the identical
eigenvalues are simply
degenerate, or whether there is a coalescence of two or even three eigenstates.
In the first case we encounter a usual EP2 and in the second an EP3. Note that
$s^{-1}$ does not exist if $s$ does not have full rank whereas $ham$ is still
well defined but may no longer be diagonalisable; this is, apart from the
branch point behaviour, the algebraic signature of an EP: instead of the
diagonal form for $ham$ there exists always the (non-diagonal) Jordan normal
form.

A judicious choice fulfilling the requirements is given by
\beq
s=\pmatrix{1 & \sqrt{1-\gamma^2} & \sqrt{1-\gamma^2} \cr
1 &  g/2-\gamma \sqrt{\frac {1-\gamma^2-g^2/4}{\gamma^2+g^2/4}}
&  g/2+\gamma \sqrt{\frac {1-\gamma^2-g^2/4}{\gamma^2+g^2/4}} \cr
1 &  (\sqrt{g/2}-\gamma \sqrt{\frac {1-\gamma^2-g^2/4}{\gamma^2+g^2/4}})^2
&  (\sqrt{g/2}+\gamma \sqrt{\frac {1-\gamma^2-g^2/4}{\gamma^2+g^2/4}})^2 }.
\label{sim}
\eeq

For $\gamma \ne \sqrt{1-g^2/4}$  one sees that encircling the critical
point in the complex plane swaps the levels $E_3$ and $E_4$ including their
(unnormalised) eigenfunctions listed in $s$ from Eq.~(\ref{sim})
$$
\pmatrix{ \sqrt{1-\gamma^2} \cr g/2-\gamma \sqrt{\frac {1-\gamma^2-g^2/4}
    {\gamma^2+g^2/4}} \cr
  (\sqrt{g/2}-\gamma \sqrt{\frac {1-\gamma^2-g^2/4}{\gamma^2+g^2/4}})^2}
\quad {\rm and} \quad
\pmatrix{ \sqrt{1-\gamma^2} \cr g/2+\gamma \sqrt{\frac {1-\gamma^2-g^2/4}
    {\gamma^2+g^2/4}} \cr
  (\sqrt{g/2}+\gamma \sqrt{\frac {1-\gamma^2-g^2/4}{\gamma^2+g^2/4}})^2}
$$
being just the characteristic of an EP2. In fact, the square root
$\sqrt{1-\gamma^2-g^2/4}$ simply changes sign when a circle is traced out in the
complex $\gamma$-plane around the point $\sqrt{1-g^2/4}$ thereby interchanging
$E_3$ and $E_4$ and their corresponding eigenvectors. Note, however, that
{\em at the critical point} all three eigenvectors also become equal (up to a
constant) thus indicating the special nature of this point. In fact,
at the critical point the rank drop of $s$ is 2, i.e.~rank($s$)=1.
At this point $ham$ cannot be diagonalised (recall $s^{-1}$ does not exist), 
but the Jordan normal form of $ham$ always exists and is given by
\beq
J[ham]=\pmatrix{g/2 & 1 & 0 \cr 0 & g/2 & 1 \cr 0 & 0 & g/2 }. \label{jord}
\eeq
This is the clear algebraic signature of an EP3, where three levels are
coalescing. Its analytic counterpart, that is the cubic root behaviour, has
been demonstrated in Section 2 by numerical means.

We emphasise that the clear signature of an EP3 as presented above is only
available when analytic expressions are at hand. To obtain this result directly
by numerical means is virtually impossible. One rather has to resort
to the means as presented in the previous section, where the existence of the
EP3 is established by invoking an asymmetry and then use closed contours in a
suitable complex parameter plane. To elucidate this pattern we discuss in the
appendix a three-dimensional matrix demonstrating the underlying mechanism; as
our matrix $ham$ is unsuitable for demonstration due to its complicated
structure we use a much simplified matrix.

\subsection{Behaviour for $g \to 0$}

All these statements hold for $g>0$ but care must be taken when $g \to 0$.
Recall that the GP-equation has a non-uniform limit for 
$g \to 0$ as is discussed in Section 2. The same holds for the matrix model
when $g \to 0$. As the limiting behaviour is subtle and difficult to obtain
numerically, we here give some results obtained in the matrix model. Note in
particular how the triple point approaches the branch point where the
$\mathcal{PT}$-eigenfunctions meet when $g \to 0$  for the matrix model
(figure~\ref{fig:spectra}) and $g \to 0$  for the numerical solution
(figure~\ref{fig:small_g}).

The eigenvalues $E_3$ and $E_4$ have, for $g>0$, the value $g/2$ at $\gamma=0$.
Yet these eigenvalues are $\pm 1$ for $g=0$ and $\gamma=0$. Similarly, the
derivative tends to infinity at $\gamma=0$ for $g \to 0$, as can be checked
analytically, and is visualized also in figure~\ref{fig:spectra}. But it is
zero when $g=0$ is taken from the outset. In other words, $E_4$ is getting
nearer and nearer to $E_2$, but at $\gamma=0$ in a non-continuous fashion
($ E_3$ approaches $-E_4$ for $g \to 0$). Note that this behaviour nicely
reflects the behaviour of the numerical solutions of the full GP-equation
(see figure~\ref{fig:small_g}).

At $\gamma=1$ it is the eigenfunctions that show the typical non-uniform
behaviour. Note that the eigenvectors as listed in Eq.~(\ref{sim}) cannot be
used in the limits $\gamma \to 1$ and $g \to 0$, irrespective of the order of
the limits taken, since $s^{-1}$ does not exist. Rather, the full matrix $ham$
must be considered. If $g=0$, corresponding to switching off the non-linearity
in the GP-equation, the eigenvectors of the two positive solutions ($E_2=E_4=
\sqrt{1-\gamma^2}$) are given by 
$$
\pmatrix{0  \cr 0  \cr 1 } \quad {\rm and} \quad \pmatrix{1  \cr 1  \cr 0 } 
$$
whereas, when the limits are commuted ($\gamma=1$ first and then $g \to 0$) the
corresponding eigenvectors are
$$
\pmatrix{0  \cr 1+\rmi  \cr 1 } \quad {\rm and} \quad
\pmatrix{0  \cr 1-\rmi  \cr 1 }  . 
$$
Recall that $E_3$ and $E_4$ are complex when $\gamma$ is to the right of the
critical point, that is if $\gamma>\sqrt{1-g^2/4}$, a condition met for the
order of the limits taken. This is reflected here in the complex eigenvectors,
even though the eigenvalues are zero in either limit.

\begin{figure}[tb]
  \begin{center}
    \includegraphics[width=\textwidth]{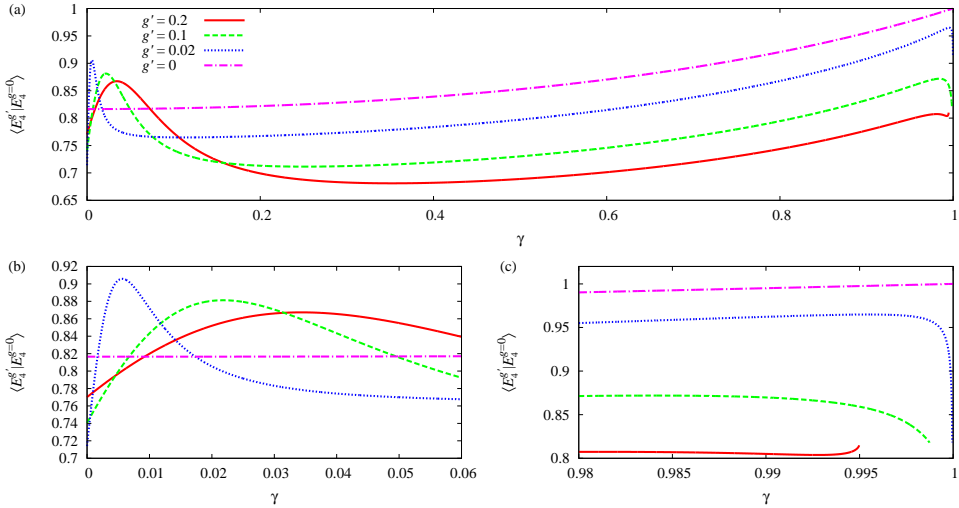}
  \end{center}
  \caption{\label{fig:scalprod} 
    Scalar product of the eigenvector for $g=0$ with the eigenvector for
    $g' \neq 0$ as function of $\gamma$ in the range $0 \le \gamma \le 1$ (a).
    The values displayed refer to $g'=0.2$ (red solid line), $g'=0.1$ (green
    dashed line), $g'=0.02$ (blue dotted line) and $g' \to 0$ (purple
    dashed-dotted line). For better illustration the $\gamma$-axis is stretched
    at $\gamma=0$ (b) and at $\gamma=1$ (c). All lines are purely real only for
    $\gamma<\sqrt{1-g^2/4}$; it explains the ending for $\gamma<1$ as only pure
    real values are displayed.}
\end{figure}

A further point of interest is the behaviour of the scalar product taken from
the eigenvectors referring to the plain solution for $g=0$ and the eigenvector
referring to $g > 0$. Of interest is the limit $g \to 0$ of the scalar
product. For the eigenvalue $E_4=g/2+\gamma \sqrt{\frac {1-\gamma^2-g^2/4}
  {\gamma^2+g^2/4}}$ we obtain for the normalised eigenvectors the scalar
product
\begin{equation}
\fl
\langle E_4^{g>0}|E_4^{g=0}\rangle  = 
\frac{g/2+ \sqrt{1-\gamma^2}+\gamma\sqrt{ \frac{1-g^2/4-\gamma^2}
    {g^2/4+\gamma^2}}} 
{\sqrt{2}\sqrt{1-\gamma^2+ (\sqrt{g/2}+\gamma \sqrt{\frac {1-\gamma^2-g^2/4}
      {\gamma^2+g^2/4}})^4 
    + (g/2+\gamma \sqrt{\frac {1-\gamma^2-g^2/4}{\gamma^2+g^2/4}})^2}}.
\label{scal}
\end{equation}
We simply list some major facts as they can be easily verified from
Eq.~(\ref{scal}). The scalar product behaves in its limit $g \to 0$
non-uniformly at both ends of the $\gamma$-interval. At $\gamma=0$
Eq.~(\ref{scal})
assumes the value $(1+g/2)/\sqrt{2+4g^2/4}$ whereas the value $\sqrt{2/3}$ is
assumed when $g = 0$ is set first. At the right end of the $\gamma$-interval the
value is unity when $g = 0$  while the value is $\sqrt{(1/2+\rmi)}\,(3-\rmi)/5$
when $\gamma =1$ is set before $g \to 0$. We mention that the intricacy of the
expansion of $\langle E_4^{g>0}|E_4^{g=0}\rangle$ in powers
of $g$ probably hints to the difficulty in approaching the
limit $g \to 0$ numerically using the GP-equation directly. In fact the
expansion reads 
$$
\frac{\sqrt{1-\gamma^2}\sqrt{2}}{\sqrt{3-4\gamma^2+\gamma^4}}-\sqrt{g}
\frac{2(\gamma^2-1)}{(\gamma^2-3)\sqrt{3-4\gamma^2+\gamma^4}}+\ldots
$$
where the $\gamma$-dependent coefficient of $\sqrt{g}$ vanishes for $\gamma=1$;
it means that the behaviour of the square root in $g$ will show for $\gamma<1$
but not for $\gamma=1$. In figure~\ref{fig:scalprod}
the scalar product is illustrated for a few values of $g$. The non-uniform
behaviour at both ends of the $\gamma$-interval is clearly visible. We note that
the scalar product assumes the value $\sqrt{2/3}$ at the critical point,
independent of $g$.

\section{Summary and Conclusions}
A number of novel aspects originating from the non-linear term $g|\psi|^2$ in
the interaction term of the Schr\"odinger equation are communicated in the
present paper. There is the occurrence of two new bound states in
the low energy spectrum. Naively, they are seen for $g>0$ only when they emerge
as complex solutions from the ground state for $\gamma \ge \gamma_{\rm cr}$
thereby breaking the $\mathcal{PT}$-symmetry of the system. 
An appropriate complex extension of the GP-equation reveals that two real
additional solutions do exist for $\gamma < \gamma_{\rm cr}$ which are in fact
eigenfunctions of $\mathcal{PT}$. A further novel result is the triple point at
$\gamma = \gamma_{\rm cr}$, in particular the singular spectral behaviour.
With a superficial look the singularity appears as a ``common'' EP2 where the
two new solutions turn complex while the ground state from which they emerge
seems to be unaffected. However, a more thorough analysis reveals that the
three states are nested in an EP3; in fact, for any generic perturbation this
result prevails.

While these results are obtained by careful numerical analysis the point
$g=0$ being a singular point of the non-linear equation seems to be
inaccessible by numerical means. In other words, the limit $g \to 0$ leading to
the simpler linear equation is subtle, there is a clear non-uniform behaviour
for the spectrum and eigenfunctions. To study this aspect in analytic terms a
matrix model simulating quantitatively the spectrum of the GP-equation is
presented. While the model elucidates the limit very clearly it also sheds
light on the specific behaviour of the triple point: the distinction between
the EP2 and EP3 behaviour depending on the specific approach of the singular
point becomes evident in analytic terms. The matrix model is designed to be
as simple and instructive as possible since our interest is focused upon the
explanation of the spectra. An exact quantitative agreement with the numerical
results is not intended as it will complicate the analytic discussion.

We believe that these findings are expected to have an impact in experimental
work in that specific effects can be detected related to the structure of the
model studied. It is well known that EPs have a definite physical significance
\cite{WDH2012} in a great variety of physical systems. The presence of the
triple point has already been established to lead to an instability in the
dynamical behaviour \cite{Dast13a}.

It remains a challenge to detect the signature of the EP3 in a BEC. Whether
or not the two additional $\mathcal{PT}$-symmetric states (for
$\gamma<\gamma_{\mathrm{cr}}$) merging at $\gamma=\gamma_{\mathrm{cr}}$ into the EP3
have physical significance is, at this stage, an open question.
However, it has already been shown that an experimental realisation of
a Bose-Einstein condensate in a $\mathcal{PT}$-symmetric double-well setup
can be achieved by embedding the double well in a partially tilted multi-well
structure \cite{Kreibich2013a}. These systems are well accessible with
today's experimental techniques \cite{Salger2007a,Foelling2007a,Henderson2009a}.
If they are exposed to loss effects they exhibit resonances, which
show as functions of the physical parameters changes of their crossing
scenarios, i.e.\ an effect typical for the presence of exceptional points
\cite{Witthaut2007a}. It will be interesting to investigate the relation
of these exceptional points with those appearing for the balanced gain
and loss scenario studied in this article.

As an alternative method for an experimental realisation one may think of
introducing spatially separated regions with gain and loss of atoms.
Bidirectional couplings between spatially separated condensates have already
been realised \cite{Shin05,Gati06}. Furthermore, electron beams have
successfully been used to introduce loss in single sites of optical lattices
\cite{Gericke08a}, and an influx of atoms can be achieved by exploiting
different electronic states of the atoms to transfer them from a reservoir
into one well of a trap \cite{Robins2008a}.

\ack
Work on this topic began at the European Centre for Theoretical Studies in
Nuclear Physics and Related Areas (ECT*), Trento, Italy. We thank the Centre
and its then director Achim Richter for their hospitality and generous support.
WDH and GW also gratefully acknowledge the support from the National Institute
for Theoretical Physics (NITheP), Western Cape, South Africa. GW expresses his
gratitude to the Department of Physics of the University of Stellenbosch for
kind hospitality while this manuscript was prepared.
\\

\section*{Appendix}
The matrix
$$
\pmatrix{2 & -1 & 0 \cr  2+y & -1 &  -y \cr -1 & 0  & 2}
$$
has the spectrum
\bea
E_1&=& 1  \nonumber \\
E_2&=& 1+\sqrt{1-y}  \nonumber  \\
E_3&=& 1-\sqrt{1-y}. \nonumber 
\eea
Clearly, and from the structure of the associated eigenfunctions, there is an
EP2 at $y=1$. However, when $y=1$ is chosen from the outset the Jordan
normal form reads
$$
\pmatrix{1 & 1 & 0 \cr 0 & 1 & 1 \cr 0 & 0 & 1 }
$$
which clearly indicates an EP3. How to find this by numerical means?

We use a perturbation $\ep$ 
$$
\pmatrix{2+\ep & -1 & 0 \cr  2+y & -1 &  -y \cr -1 & 0  & 2}.
$$
The expansions of the eigenvalues read
\bea
E_1&=& 1+\frac{2}{1-y}\ep +O (\ep^2 ) \nonumber \\
E_2&=& 1+\sqrt{1-y} + \left (\frac{1+y}{2(1-y)}+\frac{1}{2\sqrt{1-y}}
\right ) \ep +O (\ep^2 )\nonumber  \\
E_3&=& 1-\sqrt{1-y} + \left (\frac{1+y}{2(1-y)}-\frac{1}{2\sqrt{1-y}} 
\right ) \ep +O (\ep^2 ). \nonumber 
\eea
Obviously, the expansion breaks down at the critical point $y=1$.
When $y$ is set equal to $1$ in the perturbed matrix, we now see clearly how
the eigenvalues sprout from the EP3 in the expected manner. The expansions read
now
\bea
E_1&=& 1 -2^{1/3} \exp (2 \rmi \pi/3) \ep^{1/3} + O (\ep^{2/3} )  \nonumber \\
E_2&=&  1-2^{1/3} \ep^{1/3} + O (\ep^{2/3} )   \nonumber \\
E_3&=&  1-2^{1/3} \exp (-2 \rmi \pi/3) \ep^{1/3} + O (\ep^{2/3} ).  \nonumber
\eea
This pattern is basically unchanged for any generic perturbation. It clearly
demonstrates the principle underlying in the numerical manifestation of the
EP3.

\end{document}